\documentclass[12pt]{article}

\usepackage{amsmath,amsfonts,latexsym,amssymb,amscd}
\usepackage{pslatex}
\usepackage[latin1]{inputenc}
\usepackage[T1]{fontenc}
\usepackage{pspicture}
\usepackage{verbatim,amsthm,curves,graphics}
\usepackage{mathrsfs}

\usepackage{graphicx}

\newcommand{\G}{{\mathcal G}}
\newcommand{\mr}{{\mathbb R}}

\newcommand{\mc}{{\mathbb C}}
\newcommand{\mz}{{\mathbb Z}}

\begin{document}

%\begin{flushright}
%Sofia University\\
%\end{flushright}
%%%%%%%%%%%%%%%%%%%%%%%%%%%%%%%%%%%%%%%%%%%%%%%%%%%%%%%%%%%%%%%%%%%

\title{A Uniqueness theorem for  black holes with Kaluza-Klein asymptotic  in 5D Einstein-Maxwell gravity}

\author{
Stoytcho Yazadjiev$^{}$\thanks{\tt yazad@phys.uni-sofia.bg}
\\ \\
{\it $ $Department of Theoretical Physics, Faculty of Physics, Sofia
University} \\
{\it 5 J. Bourchier Blvd., Sofia 1164, Bulgaria} \\
    }
\date{}

\maketitle

\begin{abstract}
In the present paper we prove a uniqueness theorem for stationary multi black hole configurations with
Kaluza-Klein asymptotic in a certain sector of 5D Einstein-Maxwell gravity. We show that such multi black
hole configurations  are uniquely specified by the interval structure,  angular momenta of the horizons,
magnetic charges and the magnetic flux. A straightforward generalization of the uniqueness theorem for
5D Einstein-Maxwell-dilaton gravity is also given.
\end{abstract}

%%%%%%%%%%%%%%%%%%%%%%%%%%%%%%%%%%%%%%%%%%%%%%%%%%%%%%%%%%%%%%%%%%%

%\draft
\sloppy

\section{Introduction}
In the last decade we have witnessed a remarkable advent of the higher dimensional gravity
and  especially of higher dimensional black holes. Many interesting black hole solutions with amazing properties  were discovered
in various gravity theories in higher dimensional spacetimes. The accumulation of black hole solutions naturally raises the question
of their classification. The general classification for arbitrary spacetime dimensions and for all known gravity theories is formidable
task which will probably need the efforts of generations. However, in some cases the full classification is possible.
In $n=4$ spacetime dimensions, asymptotically flat, stationary vacuum or electrovac
black hole solutions in Einstein gravity are completely characterized by their asymptotic charges---mass, angular
momentum, and electric (or magnetic) charge~\cite{Israel67,Carter,Robinson,Mazur,Bunting, MAKNP, Amsel08} (see also \cite{Heusler}).
The asymptotically flat, static, vacuum and electro-vacuum black holes in
arbitrary dimensions were classified in \cite{Gibbons,Rogatko}.
The complete classification of stationary black holes in more than $n=4$ spacetime dimensions
is at present an open problem. However, in a recent paper \cite{HY1}, a partial
classification was achieved for asymptotically flat, vacuum (non-extremal) black hole solutions under the assumption that the number of
commuting axial Killing fields is sufficiently large. The particular case considered there was
$n=5$, and the number of axial Killing fields required was two.
Under this assumption, it was shown how to construct from the given solution a certain set of invariants
consisting of a set of real numbers and a collection of integer-valued vectors.
These data were called the "interval structure" of the solution. It
determines in particular the horizon topology, which could be either
$S^3, S^1 \times S^2$ or a Lens-space $L(p,q)$. It was then demonstrated that the interval structure together
with the asymptotic charges gives a complete set of invariants of the solutions, i.e.,
if these data coincide for two given solutions, then the solutions are isometric.
The generalization of \cite{HY1} for a certain sector of 5D Einstein-Maxwell gravity was done in
\cite{HY2}. In the sector under consideration, the 5D asymptotically flat Einstein-Maxwell black holes are classified in terms of their
interval structure, angular momentum and the magnetic charges associated with the generators of $H_2(M)$. Uniqueness theorems were also proven
for certain cases of the  five-dimensional minimal supergravity \cite{TYI1}-\cite{AH}. The uniqueness of the 5D extremal vacuum  black holes was considered in \cite{FL}.

Fortunately, a full classification can also be achieved for Kaluza-Klein black holes in the higher dimensional
Einstein gravity \cite{HY3}. The vacuum Kaluza-Klein black holes are again fully classified in terms of their interval structure and angular momenta.

In this paper, we generalize the analysis of \cite{HY3} to include Maxwell field.
More precisely we generalize \cite{HY3} to a certain, completely integrable sector of 5D Einstein-Maxwell gravity.
We restrict ourselves to 5 dimensions
where we are free from technical complications and where we can demonstrate the main idea in pure form.
The naive expectation is that the generalization  of uniqueness theorem for Einstein-Maxwell black holes with Kaluza-Klein asymptotic,
can be done along the lines of the similar generalization
in the asymptotically flat case, in other words in terms of the interval structure, angular momenta and the magnetic charges.
However, this is not the case. It was shown in \cite{YN1} (see also \cite{YN2}) that, in the general case,
the interval structure and the local and asymptotic charges are insufficient to
fully classify the Einstein-Maxwell black holes with Kaluza-Klein asymptotic. The very uniqueness theorem was formulated in \cite{YN1}
which states that the Kaluza-Klein black holes in  Einstein-Maxwell gravity are fully specified by their interval structure, global and local charges,
angular momenta and magnetic fluxes. The novelty in comparison with the asymptotically flat case is
the appearance of the magnetic fluxes in the conditions of the theorem.\footnote{It is  worth mentioning that, in accordance with the general 
statement of \cite{YN1}, we also need to specify the  fluxes in order to have uniqueness theorem for asymptotically flat 
multi black hole configurations in 5D minimal supergravity \cite{AH}.  } Here we give mathematically
more precise version of this theorem and its proof in 5 dimensions. The mathematical technique of
the proof is the same with that  in the asymptotically flat case \cite{HY2} which requires $\sigma$-model presentation (on a symmetric space) of the dimensionally reduced field equations. In order to have symmetric space $\sigma$-model form of the dimensionally reduced equations,
as in the asymptotically flat case, certain  additional conditions are imposed upon the Maxwell field
and the axial Killing fields. The extra assumptions placed upon
the Killing fields imply that the electric charges, and
some of the angular momenta of the horizons  vanish. They also imply that the possible interval structures are
limited. In particular, the horizons topologies can only be either $S^3$ or $S^2 \times S^1$.

Non-trivial Einstein-Maxwell black holes with Kaluza-Klein asymptotic  satisfying our assumption have been
found by~\cite{YN1,YN2}.

The paper is organized as follows. In the section 1 following \cite{HY3}, for completeness we give in concise form the necessary mathematical
base. In section 2 we consider the extra assumptions imposed on the Killing fields and the Maxwell field,
the consequences of them and the dimensionally reduced Einstein-Maxwell equations. The main result is presented in section 3.
In the Discussion we comment on possible generalizations and some tricky cases.

\section{Stationary Einstein-Maxwell black holes in $5$ dimensions}

Let $(M,g,F)$ be a $5$-dimensional, analytic,
stationary black hole spacetime satisfying the Einstein-Maxwell
equations
\begin{eqnarray}
&&R_{ab}= \frac{1}{2}\left(F_{ac}F_{b}{}^c- \frac{g_{ab}}{6}F_{cd}F^{cd} \right),\\
&&\nabla_{a}F^{ab}=0=\nabla_{[a} F_{bc]}.
\end{eqnarray}
Let $\xi$ be the
asymptotically timelike Killing field, $\pounds_\xi g = 0$,
which we assume is normalized
so that $\lim g(\xi,\xi)  = -1$ near infinity. We assume also that
the Maxwell tensor is invariant under $\xi$, in the sense that $\pounds_\xi F = 0$.
We consider 5-dimensional spacetime $M$ that has four asymptotically flat large dimensions and one asymptotically small extra dimension.
More precisely we consider 5-dimensional spacetime with asymptotic region $M_{\infty}=\mr^{3,1}\times S^1$ and asymptotic metric

\begin{eqnarray}
g= - dt^2 + dx^2_1 + dx^2_2 + dx^2_3 + d\phi^2 + O(r^{-1})
\end{eqnarray}
where $x_i$ are the standard Cartesian coordinates on $\mr^3$, $\phi$ is the standard periodic coordinate on $S^1$ with a period $2\pi$.
Here $O(r^{-1})$ stands for all metric components that drop off at least as $r^{-1}$ in the radial coordinate $r=\sqrt{x^2_1 + x^2_2 + x^2_3}$.

The domain of outer communications is defined by

\begin{eqnarray}
<<M>>={\cal I}^{+}(M_{\infty})\bigcap {\cal I}^{-}(M_{\infty})
\end{eqnarray}
where ${\cal I}^{\pm}(M_{\infty})$ denote the chronological future/past of $M_{\infty}$.

Here we will assume the existence of $2$  further axial Killing  fields $\zeta$ and $\eta$ which are mutually commuting and commute
with the asymptotically timelike Killing field $\xi$, have periodic orbits with period $2\pi$ and leave the Maxwell tensor $F$ invariant, i.e.
$\pounds_\zeta F =\pounds_\eta F = 0$. We also assume that the Killing filed $\eta$ is associated with the compact
dimension and that in the asymptotic region $M_{\infty}$ the Killing fields  $\zeta$ and $\eta$ take the standard form

\begin{eqnarray}
&&\zeta=x_1{\partial/\partial x_2} -  x_2{\partial/\partial x_1},\\
&&\eta = {\partial /\partial \phi}.
\end{eqnarray}

The group of isometries is hence ${\cal G}=\mr \times U(1)^2$, where $\mr$ stands for the flow of $\xi$ while $U(1)^2$
corresponds to the commuting flows of axial Killing fields.

As a part of our technical assumptions we further assume that \cite{HY3}:
\medskip
\noindent

a) $<<M>>$ contains an acausal, spacelike, connected hypersurface $\Sigma$ asymptotic to a $t=const$ slice in the asymptotic
region $M_{\infty}$, whose closure has as its boundary $\partial \Sigma=\bigcup_i {\cal H}_i$ cross sections of the horizons.
\medskip
\noindent

b) The horizon cross sections are compact and the horizons are non-degenerate.
\medskip
\noindent

c) The orbits of the Killing field $\xi$ are complete.

\medskip
\noindent

d) $<<M>>$ is globally hyperbolic.

\medskip
\noindent

Due to the symmetries of the spacetime the natural space to work on is the orbit (factor) space ${\hat M}=<<M>>/\G$,
where $\G$ is the isometry group. The structure of the factor space is described by the following theorem which is a
straightforward generalization of the corresponding theorem in \cite{HY3}:

\medskip
\noindent
{\bf Theorem:} Let $(M,g)$ be a stationary, asymptotically Kaluza-Klein, 5-dimensional black hole spacetime with isometry group
$\G=\mr\times U(1)^2$ satisfying the technical assumptions stated above. Then the orbit space ${\hat M}=<<M>>/\G$ is
a 2-dimensional manifold with boundaries and corners homeomorphic to a half-plane. Some boundary segments $I_i\subset \partial {\hat M}$
correspond to the quotients of the horizons ${\cal H}_i=H_i/\G$, while the remaining segments $I_j$ correspond to the various axes, where
a linear combination $a_\zeta(I_j) \zeta + a_\eta (I_j)\eta=0$ and  ${\bf a}(I_j)=(a_\zeta(I_j),a_\eta(I_j))\in \mz^2$.  For adjacent intervals $I_j$
and $I_{j+1}$ (not including the horizons) the vectors ${\bf a}(I)=(a_\zeta(I),a_\eta(I))$  are subject to the following constraint
\begin{eqnarray}\label{constraint}
|\det \left(
       \begin{array}{cc}
         a_\zeta(I_j) &  a_\zeta(I_{j+1})\\
          a_\eta(I_j)& a_\eta(I_{j+1}) \\
       \end{array}
     \right)|=1 .
\end{eqnarray}
\medskip
\noindent

In the interior of ${\hat M}$ there is a naturally induced metric ${\hat g}$ which has signature $++$. We denote derivative operator associated with
${\hat g}$ by ${\hat D}$. Let us now consider the Gramm matrix of the Killing fields $G_{IJ}=g(K_{I},K_{J})$,
where $K_{0}=\xi$, $K_{1}=\zeta$ and $K_{3}=\eta$. Then the determinant $\rho^2=|\det G|$ defines a scalar
function $\rho$ on ${\hat M}$ which, as  well known, is harmonic,
${\hat D}^{a}{\hat D}_{a}\rho=0$ as a consequence of the Einstein-Maxwell field equations.
It can be shown that $\rho>0$, ${\hat D}_a \rho\ne 0$ in the interior of ${\hat M}$ and that $\rho=0$ on $\partial {\hat M}$.
We may define a conjugate harmonic function $z$ on ${\hat M}$ by $dz={\hat \star}\, d\rho$, where ${\hat \star}$ is the Hodge dual on ${\hat M}$.
The functions $\rho$ and $z$ define global coordinates on ${\hat M}$
identifying the orbit space with the upper complex half-plane
\begin{eqnarray}
{\hat M}= \{z+ i\rho \in \mc, \rho\ge 0  \}
\end{eqnarray}
with the boundary corresponding to the real axis. The induced metric ${\hat g}$ is given in these coordinates by

\begin{eqnarray}
{\hat g}= \Omega^2(\rho,z)(d\rho^2 + dz^2),
\end{eqnarray}
$\Omega^2$ being a conformal factor.

The above theorem allows us to introduce the notion of {\it interval structure}. The orbit space of the domain of outer communication
by the isometry group is a half plane ${\hat M}=\{z+i\rho, \rho>0\}$ and its boundary $\partial {\hat M}$ is divided into a finite number of
intervals $I_j$:

\begin{eqnarray}
(-\infty,z_1), (z_1,z_2),...,(z_N,z_{N+1}),(z_{N+1},+\infty) .
\end{eqnarray}

 To each interval we associate its length $l(I_j)$ and a vector ${\bf a}(I_j)=(a_{\zeta}(I_j),a_\eta(I_j))\in \mz^2$ (subject to (\ref{constraint}))
when the interval does not correspond to a horizon. To intervals corresponding to the orbit spaces ${\cal H}_i$ of the horizons we associate
zero vector $(0,0)$. \emph{The data ${l(I_j)}$ together with  ${\bf a}(I_j)=(a_\zeta(I_j), a_\eta(I_j))$  are called interval structure.} The vectors
${\bf a}(I_j)= (a_{\zeta}(I_j),a_\eta(I_j))$ corresponding to the outermost intervals $(-\infty,z_1)$ and $(z_{N+1},+\infty)$ must be $(1,0)$ and $(1,0)$ since the spacetime is asymptotically Kaluza-Klein.

Furthermore, we have the following theorem about the topology of the horizons \cite{HY1,HY2,HY3}

\medskip
\noindent
{\bf Theorem:} Under the assumptions made above the horizon cross sections ${\cal H}_i$ must be topologically
either $S^2\times S^1$, $S^3$ or a Lens space $L(p,q)$ ($p,q \in \mz$). Here $p$ is given by  $p=\det({\bf a}_{h_i-1}, {\bf a}_{h_i+1})$
where ${\bf a}_{h_i-1}$ and ${\bf a}_{h_i+1}$ are vectors adjacent on the $i$-th horizon ${\cal H}_i$. The topology of ${\cal H}_i$
is $S^2\times S^1$ for $p=0$, $S^3$ for $p=\pm 1$ and $L(p,q)$ in the other cases.
\medskip
\noindent

\section{Dimensionally reduced Einstein-Maxwell equations, magnetic charges and magnetic flux}

In the present paper we will not consider the most general 5D Einstein-Maxwell black holes.
We will focus ourselves to black holes in a certain sector of 5D Einstein-Maxwell gravity
which is known to be completely integrable \cite{Yazadjiev}. The simplifying assumptions we make in addition to the general
hypothesis stated above are the following:

\medskip
\noindent
1) We assume that the  axial Killing field $\eta$ is  orthogonal to the other Killing fields, $g(\zeta,\eta)=g(\xi,\eta)=0$,
and that it is  also hypersurface orthogonal, $\eta \wedge d\eta=0 $.

\medskip
\noindent
2) About the Maxwell 2-form $F$ we assume that the following conditions are satisfied

\begin{eqnarray}
i_\xi F=i_\zeta F=i_\eta \star F=0 .
\end{eqnarray}

Let us consider the consequences of the assumptions 1) and 2). The first consequence of 1) is  that the angular momentum
associated with $\eta$ of every horizon ${\cal H}_i$,
defined\footnote{In the present paper the angular momenta are defined up to irrelevant numerical factor.} by

\begin{eqnarray}
J^{\,i}_{\eta}=\int_{{\cal H}_{i}} \star d\eta
\end{eqnarray}
is zero, $J^{\,i}_{\eta}=0$. Secondly, since the Killing vector $\eta$ is orthogonal to $\zeta$, if at some spacetime point we have
$a_\zeta \zeta + a_\eta \eta=0$, then either $(a_\zeta,a_\eta)=(0,0)$ or $(a_\zeta,a_\eta)=(1,0)$, $(a_\zeta,a_\eta)=(0,1)$, or both axial
Killing fields $\zeta$ and $\eta$ vanish. Thus the assumption 1) restricts the possible interval structures. However, the known
exact solutions fall in these restricted interval structures.
In turn, the possible topologies of the horizons are also restricted and they are either $S^2\times S^1$ or $S^3$.
This immediately follows from the theorem about the  topologies of the horizons.

Now let us consider the consequences of the assumption 2). From $i_\xi F=0$ it follows that the electric charge of every horizon ${\cal H}_i$,
defined by

\begin{eqnarray}
q^{\,i}=\int_{{\cal H}_i} \star F
\end{eqnarray}
is zero, $q^{\,i}=0$. Furthermore, all the equations in assumption 2) show that the Maxwell field is completely characterized by the 1-form
$i_\eta F$. It is easy to see that this form is closed. Indeed we have $di_\eta F=\pounds_{\eta}F - i_\eta dF=0$.

Proceeding further we define the twist 1-form by

\begin{eqnarray}
\omega = \star (\zeta\wedge\eta\wedge d\zeta )=i_\eta i_\zeta \star d\zeta.
\end{eqnarray}
Using the equations of assumption 2) and the fact that the Killing fields commute, one can show that the twist 1-form is closed, $d\omega=0$.

Both 1-forms $\omega$ and $f=i_\eta F$ are invariant under the spacetime symmetries and therefore they induce corresponding 1-forms ${\hat \omega}$
and ${\hat f}$ on the orbit space ${\hat M}$, which are still closed. Since the orbit space ${\hat M}$ is simply connected, there exist globally
defined potentials $\chi$ and $\lambda$ such that ${\hat \omega}=d\chi$ and ${\hat f}=d\lambda$ on ${\hat M}$. The potential $\lambda$ and $\chi$
play important role in writing down the dimensionally reduced Einstein-Maxwell equations on the orbit space. Let $u$, $w$ and $\Gamma$ be functions
on ${\hat M}$ defined by

\begin{eqnarray}
e^{2u}= g(\eta,\eta),\; \;\;\; e^{-u +2w}=g(\zeta,\zeta), \;\;\; \; e^{-u+ 2w + 2\Gamma}= g(\nabla \rho,\nabla \rho).
\end{eqnarray}

Then the Einstein-Maxwell equations are equivalent to the following set of
equations on the orbit space $\hat M$~\cite{Yazadjiev} :
\begin{eqnarray}\label{smodel}
\hat D^a \left(\rho \Phi^{-1}_{1} \hat D_{a} \Phi_{1}^{}\right)&=&0 \, ,\nonumber\\
\hat D^a \left(\rho \Phi^{-1}_{2} \hat D_{a} \Phi_{2}^{}\right)&=&0 \, ,
\end{eqnarray}
together with
\begin{eqnarray}\label{nudef}
-\rho^{-1} (\hat D^a \rho) \hat D_a \Gamma &=&
\left[
{3\over 8} {\rm Tr} \left(\hat D^a \Phi_{1}^{} \hat D^b \Phi^{-1}_{1}\right)+
{1\over 8} {\rm Tr} \left(\hat D^a \Phi_{2}^{} \hat D^b \Phi^{-1}_{2}\right)
\right] \,
\left[ \hat g_{ab} - 2(\hat D_a z) \hat D_b z \right] \nonumber \\
-\rho^{-1} (\hat D^a \rho) \hat D_a \Gamma &=&
\left[
{3\over 4} {\rm Tr}\left( \hat D^a \Phi_{1}^{} \hat D^b \Phi^{-1}_{1} \right)+
{1\over 4} {\rm Tr}\left( \hat D^a \Phi_{2}^{} \hat D^b \Phi^{-1}_{2} \right)
\right] (\hat D_a \rho) \hat D_b z \, ,
\end{eqnarray}
where the matrix fields are defined in terms of $u, w,\lambda,\chi$ by
\begin{eqnarray}
\Phi_{1} = \left(%
\begin{array}{cc}
  e^{u} + {1\over 3}\lambda^2 e^{-u} & {1\over \sqrt{3}} \lambda e^{-u} \\
 {1\over \sqrt{3}} \lambda  e^{-u}& e^{-u} \\\end{array}%
\right) ,
\end{eqnarray}
and
\begin{eqnarray}
\Phi_{2} = \left(%
\begin{array}{cc}
  e^{2w} + 4\chi^2e^{-2w} & 2\chi e^{-2w} \\
 2\chi e^{-2w} & e^{-2w} \\\end{array}%
\right).
\end{eqnarray}
The first two equations state that each of
the matrix fields $\Phi_{1}$ and $\Phi_{2}$ satisfies the equations of a 2-dimensional
sigma-model. The matrix fields are real, symmetric, with determinant equal to $1$ on the interior of $\hat M$.
The equations~\eqref{nudef} are decoupled from the
sigma-model equations and determine the function $\Gamma$.

Before closing this section we shall introduce the magnetic charges and the magnetic flux associated with the interval structure.
The magnetic charges are defined by

\begin{eqnarray}
 Q[C_k]= \int_{C_k}F
\end{eqnarray}
where $C_k$, $k=1,2,...$ are all the topologically inequivalent, non-contractible, closed 2-surfaces in the domain of outer communications.
The explicit construction of $C_k$ is as follows \cite{HY2}. We consider all possible curves $\hat \gamma_k, k=1,2,\dots$ in $\hat M$ with
the property that $\hat \gamma_k$ starts
on an interval labeled $(0,1)$, and ends on another interval labeled $(0,1)$, with no interval with label $(0,1)$ in
between. If we now lift $\hat \gamma_k$ to a curve $\gamma_k$ in $<<M>>$, and act with all isometries generated by
$\eta$ on the image of this curve,
then we generate a closed 2-surface $C_k$ in $<<M>>$,
which is topologically a 2-sphere for all $k$.
We may repeat this by replacing $\hat \gamma_k, k=1,2, \dots$ with a set of curves each
starting on an interval labeled $(1,0)$, and ending on another interval labeled $(1,0)$,
with no interval with label $(1,0)$ in between. If we again lift these curves to curves in $<<M>>$, and act with all isometries generated by
$\zeta$, then we generate a set of topologically inequivalent closed 2-surfaces $\tilde C_l, l=1,2, \dots$ in $<<M>>$,
each of which is topologically a 2-sphere. It may be seen that the set of 2-surfaces $\{ C_k, \tilde C_l \}$ forms a basis of $H_2(<<M>>)$.

The magnetic charges $Q[\tilde C_l]$ are not needed and in fact vanish, due to assumptions 1) and 2)  of this
section.

The magnetic flux $\Psi^{+}$ is defined by

\begin{eqnarray}
\Psi^{+}= \int_{C^{+}} F
\end{eqnarray}
where $C^{+}$ is a 2-surface with the topology of disk which is constructed as follows \cite{YN1}. Let us consider the rightmost
interval\footnote{In other words we consider the rightmost bubble.} $(z_{N-1},z_{N})$ with vector $(0,1)$ and the semi-infinite interval ${\hat \gamma}^{+}=[z_{N},+\infty)$.
We lift ${\hat \gamma}^{+}$ to a curve $\gamma^{+}$ in $<<M>>$, and act on it with the isometries generated by $\eta$. Since $\eta|_{z_N}=0$
the generated 2-surface has disc topology. In the same way we can define the magnetic flux associated with the leftmost interval with
vector $(0,1)$. However, both fluxes are not independent and
satisfy the relation $\Psi^{+}+\Psi^{-}=-2\pi\sum_k  Q[C_k]$ (see for example \cite{YN2}).

\section{Uniqueness theorem }

The central result of the present paper is given in the following theorem

\medskip
\noindent
{\bf Uniqueness Theorem:} Consider two stationary, asymptotically Kaluza-Klein, Einstein-Maxwell
black hole spacetimes of dimension 5,  having  one
time-translation Killing field and two axial Killing fields and satisfying all technical assumptions stated above.
We also assume that the Killing and
Maxwell fields satisfy the assumptions 1) and 2) above, implying that
${\bf a}(I_j) = (1,0)$ or $(0,1)$, and ${\mathcal H}_{\,i} = S^3$ or $S^1 \times S^2$, and $q^{i} = 0 = J^{i}_\eta$ for the solutions. If
the two solutions have the same interval structures, same horizon angular momenta  $J^{\,i}_\zeta$, the same
magnetic charges $Q[C_l]$ for all 2-cycles $C_l$, and same  magnetic fluxes $\Psi^{+}$, then they are isometric.
\medskip
\noindent

\medskip
\noindent

{\bf Remark:} This uniqueness theorem obviously holds also in the case when the solutions do not possess any horizon.
As an explicit example we may give the solutions describing  magnetized bubbles \cite{YN1}.
\medskip
\noindent

{\bf Proof:}
Consider two solutions $(M,g,F)$ and $(\tilde M, \tilde g, \tilde F)$
as in the statement of the theorem. We use the same "tilde" notation to distinguish any quantities associated with the two solutions.
Since the interval structures of  both solutions are the same, $<<M>>$ and $<<\tilde M>>$ can be identified as manifolds. Thus, we may assume that $<<{\tilde M}>> = <<M>>$, and that $\tilde \xi = \xi$,
$\tilde \zeta = \zeta$ and $\tilde \eta=\eta$. We may also assume that $\tilde \rho=\rho$ and $\tilde z=z$. As a consequence of these identifications, it is possible to combine the divergence identities  (\ref{smodel}) to the following Mazur identities

\begin{eqnarray}\label{Mazurid}
{\hat D}^{a}\left(\rho {\hat D}_a\sigma_{m}\right)=\rho {\hat g}^{ab}Tr\left(N^T_{m\,a}N_{m\, b}\right)
\end{eqnarray}
where $m=1,2$ and

\begin{eqnarray}
\sigma_{m}= Tr\left({\tilde \Phi}_m\Phi_{m}^{-1}- I\right),\;\;
N_{(m)a}={\tilde S}_{m}^{-1}\left({\tilde \Phi}_{m}^{-1}{\hat D}_{a}{\tilde \Phi}_{m} - {\tilde \Phi}_{m}^{-1}{\hat D}_{a}{\tilde \Phi}_{m} \right)S_{m}.
\end{eqnarray}
Here the matrices $S_{m}$ and ${\tilde S}_{m}$ are defined by $\Phi_m=S_{m}^{T}S_{m}$ and ${\tilde \Phi}_m={\tilde S}_{m}^{T}{\tilde S}_{m}$.
The key and nice point about the Mazur identities (\ref{Mazurid}) is that the right hand sides are nonnegative while the left hand sides
are total divergences.

At this stage it is convenient to view $\rho$ and $z$ as cylindrical coordinates in an auxiliary space $\mr^3$ consisting of the points $X=(\rho\cos\varphi,\rho\sin\varphi,z)$. It is also convenient to view $\sigma_m$ as rotationally symmetric functions on the
auxiliary space $\mr^3$. Then, according to the Mazur identities we have

\begin{eqnarray}
\Delta \sigma_{m}\ge 0, \; \mr^3 \backslash\{z-axis\}
\end{eqnarray}
where $\Delta$ is the ordinary Laplacian on $\mr^3$. Furthermore, the functions $\sigma_m$ are nonnegative, $\sigma_m\ge 0$.
Indeed, we have

\begin{equation}\label{sigdef1}
\sigma_1 = {\rm Tr} \Big[ \Phi_1^{-1} \tilde \Phi_1^{} - I \Big] =  \frac{(e^{u}-e^{\tilde u})^2}{e^{u}e^{\tilde u}} + \frac{1}{3}
\frac{\left(\tilde \lambda - \lambda \right)^2}{e^{u}e^{\tilde u}}\ge 0
\end{equation}
and
\begin{equation}\label{sigdef2}
\sigma_2 = {\rm Tr} \Big[ \Phi_2^{-1} \tilde \Phi_2^{} - I \Big] =  \frac{(e^{2w}-e^{2\tilde w})^2}{e^{2w}e^{2\tilde w}} + 4
\frac{\left(\tilde \chi - \chi \right)^2}{e^{2w}e^{2\tilde w}} \ge 0\, .
\end{equation}

According to the maximum principle~\cite{Weinst1,Weinst2},
if $\sigma_m$ are globally bounded above on the
entire $\mr^3$ including the $z$-axis and infinity where they vanish,
then they vanish identically. In order to show that $\sigma_m$ are bounded we must consider the behavior of  $\sigma_m$
on (i) the horizons, (ii) on the axes of $\zeta$ and $\eta$, (iii)  near infinity and (iv) on the corners.
\medskip
\noindent

(i) Obviously, on the open intervals corresponding to the horizons $\sigma_m$ are bounded. Indeed, neither $e^{u}$ nor $e^{w}$ vanish, since
both Killing fields $\zeta$ and $\eta$ are non-vanishing on the  open intervals corresponding to the horizons.
\medskip
\noindent

(ii) We first consider open intervals corresponding to $\zeta=0$ and $\eta\ne 0$, in other words intervals with vector ${\bf a}=(1,0)$.
For such intervals $e^{2u}=g(\eta,\eta)\ne 0$ and $g(\zeta,\zeta)=e^{2w-u}\to 0$ which means that $e^{2w}\to 0$. Moreover, the smoothness
of the solution requires $e^{2w}={\cal O}(\rho^2)$ near the considered intervals. From the explicit forms of the functions $\sigma_m$
it is clear that only $\sigma_2$ is potentially unbounded. The first term in $\sigma_2$ is obviously bounded. In order to show that the second
term in $\sigma_2$ is also bounded we shall consider the behavior of the twist potential near the axis \cite{HY2,HY3}. From the fact that the
twist 1-form $\omega$ vanishes on any axis by definition, the twist potential $\chi$ is constant on $z$-axis outside the intervals
corresponding to the horizons. The difference between the constant value $\chi_i$ of the twist potential on the $z$-axis left
and right to a given horizon is \cite{HY3}

\begin{eqnarray}
\chi_i(\rho=0,z_{h+1})- \chi_i(\rho=0,z_{h})=\frac{1}{(2\pi)^2}J^{\,h}_{\zeta}
\end{eqnarray}
where $J^{h}_{\zeta}$ is the angular momentum of the horizon. The same formula holds for the tilde solution

\begin{eqnarray}
{\tilde \chi}_i(\rho=0,z_{h+1})- {\tilde \chi}_i(\rho=0,z_{h})=\frac{1}{(2\pi)^2}{\tilde J}^{\,h}_{\zeta}.
\end{eqnarray}

By assumption we have $J^{\,h}_{\zeta}={\tilde J}^{\,h}_{\zeta}$ which means that ${\tilde \chi}_i-\chi_i=const$ on the $z$-axis
outside the intervals corresponding to the horizons. Since $\chi$ is defined up to a constant, we can chose this constant so that
${\tilde \chi}_i=\chi_i$ on the $z$-axis (outside the intervals
corresponding to the horizons). This together
with the fact that $\omega=d\chi$ also vanishes on the axes of $\zeta$ and $\eta$, shows that ${\tilde \chi}_i-\chi_i={\cal O}(\rho^2)$
near these axes,  which in turn implies that the second term in $\sigma_2$ is bounded.

Let us now consider the second case when $\eta=0$ and $\zeta\ne 0$, i.e. open intervals with vector ${\bf a}=(0,1)$. In the case under consideration
the smoothness of the solutions requires $e^{u}={\cal O}(\rho)$ and $e^{2w}={\cal O}(\rho)$ near the point where $\eta=0$ and $\zeta\ne 0$.
These behaviors  guarantee that $\sigma_2$ is bounded. It is also clear that the first term in $\sigma_1$ is bounded.
In order to show that the second term  in $\sigma_1$ is bounded we shall consider the behavior of the potential $\lambda$ near the points
where $\eta=0$ and $\zeta\ne 0$. Since $i_\eta F$ vanishes on the axes of $\eta$, it follows that $\lambda$ is constant on these axes.
The difference between the constant value $\lambda_i$  on two neighbor $\eta$-axes connected by the curve ${\hat \gamma}_i$ is given by

\begin{eqnarray}
\lambda_{i+1}- \lambda_{i}= \int_{{\hat \gamma}_i}d\lambda=\frac{1}{2\pi} \int_{C_i}F=\frac{1}{2\pi} Q[C_i]
\end{eqnarray}
and a similar expression for the tilde solution

\begin{eqnarray}
{\tilde \lambda}_{i+1}- {\tilde \lambda}_{i}= \frac{1}{2\pi} {\tilde Q}[C_i].
\end{eqnarray}

From these expressions and our assumption that $Q[C_i]={\tilde Q}[C_i]$, we conclude that ${\tilde \lambda}_i - \lambda_i=const$ on the axes of $\eta$.
Since $\lambda$ is defined up to a constant we can choose this constant so that  ${\tilde \lambda}_i=\lambda_i$ on the $\eta$-axes.
This, together with the fact that $d\lambda$ also vanishes on the $\eta$-axes  implies that ${\tilde \lambda} -\lambda={\cal O}(\rho^2)$
near the axes of $\eta$ and therefore the second term in $\sigma_1$ is also bounded.

\medskip
\noindent

(iii) Let  us consider the behavior of $\sigma_m$ near infinity. In order to show that $\sigma_2$ is bounded near infinity, we use that both metrics are asymptotically Kaluza-Klein and have the same
asymptotic angular momenta,  ${\tilde J}_{\zeta}=\sum_i {\tilde J}^{\,i}_{\zeta}=J_{\zeta}=\sum_i J^{\,i}_{\zeta}$.
 The fact that  $J_{\zeta}=\sum_i J^{\,i}_{\zeta}$
(and the same expression for the tilde solution) can be proven by using condition 2) and by applying Gauss theorem to the definition formula
of the asymptotic angular momentum
\begin{eqnarray}
J_{\zeta}= \int_{S^2\times S^1} \star d\zeta
\end{eqnarray}
where the integration is performed over a surface at  infinity. As a consequence of ${\tilde J}_{\zeta}=J_{\zeta}$, one can show that \cite{HY3}

\begin{eqnarray}
e^{2{\tilde w}} - e^{2w}=O(r^{-1}), \;\;\;\;\;\;
{\tilde \chi} - \chi = O(r^{-1}).
\end{eqnarray}
Hence we find  that $\sigma_{2}|_{\infty}=0$.

One can show that $\lambda$ has the following asymptotic behavior

\begin{eqnarray}
\lambda= \lambda_{\infty} + O(r^{-1})
\end{eqnarray}
where $\lambda_{\infty}$ is a constant. Now taking into account this asymptotic and the asymptotic of $e^{u}$, namely $e^u \to 1$, for   $\sigma_1$
we find

\begin{eqnarray}
\sigma_1|_{\infty}= \frac{1}{3} \left({\tilde \lambda }_{\infty} - \lambda_{\infty} \right)^2.
\end{eqnarray}

From the fact that ${\tilde \lambda}=\lambda$ on the axes of $\eta$ we can not conclude that ${\tilde \lambda}_{\infty}=\lambda_{\infty}$
 since the spacetime is asymptotically Kaluza-Klein and no axis of $\eta$ reaches infinity. At this
stage namely we must use our assumption that both solutions have the same magnetic flux. Calculating the magnetic flux we find

\begin{eqnarray}
\Psi^{+}= \int_{C^{+}}F=2\pi \int^{\infty}_{z_N}d\lambda=2\pi \left(\lambda_{\infty} -\lambda(\rho=0,z_N)\right)
\end{eqnarray}
and a similar expression for the tilde solution\footnote{In the expression of ${\tilde \Psi}^{+}$ we have taken into account that
${\tilde \lambda}(\rho=0,z_N)=\lambda(\rho=0,z_N)$.}

\begin{eqnarray}
{\tilde \Psi}^{+}=2\pi \left({\tilde \lambda}_{\infty} -\lambda(\rho=0,z_N)\right).
\end{eqnarray}
Here $z_N$ is the right boundary of the rightmost axis of $\eta$. By assumption ${\tilde \Psi}^{+}=\Psi^{+}$ which means that
${\tilde \lambda}_{\infty}=\lambda_{\infty}$. Therefore we find that $\sigma_1|_{\infty}=0$.

\medskip
\noindent

(iv) We must also consider the behavior of $\sigma_m$ at the  corners. The continuity argument  shows that $\sigma_m$ are bounded
on the corners.

\medskip
\noindent

Summarizing, we have shown that the functions $\sigma_m$, $(m=1,2)$ are bounded above on the entire $\mr^3$ including the $z$-axes and infinity, where
$\sigma_m$ vanish. Therefore, by the maximum principle \cite{Weinst1,Weinst2}, $\sigma_{m}$ vanish identically. Consequently, it immediately   follows
that ${\tilde u}=u$,   ${\tilde w}=w$, ${\tilde \chi}=\chi$, ${\tilde \lambda}=\lambda$ and ${\tilde \Gamma}=\Gamma$. From these equalities,
as in the asymptotically flat case \cite{HY2},
one can show that ${\tilde g}=g$ and ${\tilde F}=F$. This completes the proof.

\medskip
\noindent

{\bf Remark:} We can consider the other case when the Killing field $\zeta$ is hypersurface orthogonal, $\zeta\wedge d\zeta=0$, and when the electromagnetic field is along the noncompact direction $\zeta$, i.e. when the Maxwell 2-form $F$ satisfies the conditions $i_\xi F=i_\eta F=i_\zeta \star F=0$.
In this case one can show that the black hole configurations
are fully determined only in terms of the interval structure, the angular momenta\footnote{The angular momenta $J_\zeta^{\,i}$  are zero.} $J^{\,i}_{\eta}$ of the horizons and the magnetic charges $Q[{\tilde C}_k]$.

\medskip
\noindent

%%%%%%%%%%%%%%%%%%%%%%%%%%%%%%%%%%%%%%%%%%%%%%%

\section{Discussion}

Let us discuss some generalizations of our result. The proven uniqueness theorem can be  extended to the 5D Einstein-Maxwell-dilaton gravity
which can be derived from  the  Lagrangian
\begin{eqnarray}
 {\mathbf  L} = \star R - 2d\varphi \wedge \star d\varphi - {1\over 2}e^{-2\alpha\varphi} F\wedge \star F
\end{eqnarray}
where $\varphi$ is the dilaton field and $\alpha$ is the dilaton coupling parameter.

The $\sigma$-model presentation of the dimensionally reduced  5D Einstein-Maxwell-dilaton equations with the restrictions 1) and 2),
was given in \cite{Yazadjiev1}. On this base and applying the  mathematical technique of the present work,
one can prove the following uniqueness theorem

\medskip
\noindent
{\bf Uniqueness Theorem:} Consider two stationary, asymptotically Kaluza-Klein, Einstein-Maxwell-dilaton
black hole spacetimes of dimension 5,  having  one
time-translation Killing field and two axial Killing fields and satisfying all technical assumptions stated above.
We also assume that the Killing and
Maxwell fields satisfy the assumptions 1) and 2) above, implying that
${\bf a}(I_j) = (1,0)$ or $(0,1)$, and ${\mathcal H}_{\,i} = S^3$ or $S^1 \times S^2$, and $q^{i} = 0 = J^{i}_\eta$ for the solutions. If
the two solutions have the same interval structures, same horizon angular momenta  $J^{i}_\zeta$, the same
magnetic charges $Q[C_l]$ for all 2-cycles $C_l$, same value of the dilaton field at infinity $\varphi_{\infty}$
and same  magnetic fluxes $\Psi^{+}$, then they are isometric.
\medskip
\noindent

As a part of the technical assumptions in this theorem  we obviously assume that the dilaton field is invariant under the spacetime symmetries,
$\pounds_{\xi}\varphi=\pounds_{\zeta}\varphi=\pounds_{\eta}\varphi=0$.

The restrictions 1) and 2)  play very important role in the present uniqueness theorem as well as
in the uniqueness theorem for the asymptotically flat case \cite{HY2}. Due to these restrictions the target space of the potentials
is a symmetric space which ensures the complete integrability of the considered sector and  the existence of Mazur identities which are a key moment
in the proof. The natural step in generalizing the uniqueness theorems for Kaluza-Klein and asymptotically flat black holes  is to
remove the restrictions 1) and 2), in other words to consider the general case when the electromagnetic field is completely excited.
In the general case, however, it seems that the dimensionally reduced  5D Einstein-Maxwell(-dilaton) gravity  equations do not possess large
enough group of symmetries which could ensure complete integrability and the existence of Mazur identities. So, in the general case
the mathematical technique used for proving the uniqueness theorem  in the present paper and in \cite{HY2}, is not applicable.
New technique must be used in order to prove the uniqueness theorems in the general case \cite{Yazadjiev2}.

Contrary to the 5D Einstein-Maxwell(-dilaton) gravity, the 5D minimal supergravity is completely integrable in the general
case when the electromagnetic field is fully excited \cite{FJRV} (see also \cite{BCCGSW} and \cite{CBJV}). This fact shows that the uniqueness theorem of the present paper can be easily generalized within the framework of the 5D minimal supergravity, of course with the corresponding technical complications and extensions. We hope to give the formal mathematical results in a future work.

Finally, we would like to comment on following. Do the collection of the interval structure, local and asymptotic charges
(and the magnetic fluxes) and angular momenta always fully determine the black hole solutions? Fortunately or unfortunately the answer seems to be  "NO". As our preliminary
numerical calculations show there could exist many (even infinitely many) black hole solutions with the same interval structure, angular momenta
and local and asymptotic charges. Such a behavior is observed in some dilaton gravity models coupled to the
electromagnetic field with an appropriate dilaton coupling function \cite{DYS}.

\vspace{1cm}

\noindent
{\bf Acknowledgements:}
This work was partially supported by the Bulgarian National Science Fund under Grants DO 02-257, VUF-201/06 and by Sofia University Research Fund
under Grant No 074/2009.

\end{document}